# Anomalous behavior in high-pressure carbonaceous sulfur hydride

Mehmet Dogan[1,2], Marvin L. Cohen[1,2,*]

[1] Department of Physics, University of California, Berkeley, CA 94720, USA
[2] Materials Sciences Division, Lawrence Berkeley National Laboratory, Berkeley, CA 94720, USA
[*] To whom correspondence should be addressed: mlcohen@berkeley.edu

**Abstract**

A new experimental study by Snider *et al.* [2020 *Nature* **586** 373–7] reported behavior in a high-pressure carbon-sulfur-hydrogen system that has been interpreted by the authors as superconductivity at room temperature. The sudden change of electrical resistance at a critical temperature and the change of the *R vs. T* behavior with an applied magnetic field point to superconductivity. This is a very exciting study in one of the most important areas of science, hence, it is crucial for the community to investigate these findings and hopefully reproduce these results. In this paper, we present calculations that expand upon the arguments put forth by Hirsch and Marsiglio [arXiv:2010.10307], and offer some speculations about physical mechanisms that might explain the observed data. We show that there are errors in the analysis presented in the experimental paper, and with the correct analysis, the reported *R vs. T* data significantly deviate from the expected behavior. In particular, the extremely sharp change in resistance at the superconducting transition is not consistent with a strongly type II superconductor.

**Introduction**

The long-running effort to achieve superconductivity at room temperature received a boost in 2015 when a sulfur hydride ($H_3S$) was reported to have a critical temperature as high as ~200 K at high pressures [1]. Over the course of the following few years, other hydrides have followed, most notably lanthanum hydride ($LaH_{10}$), which has a critical temperature of ~260 K [2,3]. The low nuclear mass of hydrogen and the consequent high vibrational frequencies have been thought as the main cause of the high-temperature superconductivity in hydrides, which is what originally motivated these experimental studies [4,5]. Hence, an enormous amount of ongoing experimental effort is being expended toward high-pressure hydrides. In the same vein, Snider *et al.* recently reported an experimental work where a high-pressure carbonaceous sulfur hydride (C-S-H) goes through a sharp change in resistance at a critical temperature of up to ~290 K [6]. Among the reported data, the change of the transition temperature with respect to the magnetic field as well as the a.c. magnetic susceptibility measurements point to superconductivity. If this material is indeed a superconductor, this would mark the achievement of the century-long goal of obtaining a room-temperature superconductor.

In this paper, we present an analysis of the experimental results based on well-established theories of superconductivity and find that some of the observations deviate from the expected behavior of superconductors. In particular, we expand upon the arguments of Hirsch and Marsiglio (H&M) [7], and show that the material should be strongly type II, which then makes the observation of extremely sharp resistance *vs.* temperature curves anomalous. We also estimate the resistivity of the material, which falls into the typical metal range below the critical temperature. Finally, we suggest some alternative physical processes that might explain the observations, if the C-S-H system is not a superconductor.

**Methods and Results**

In the experiment by Snider *et al.* [6], a critical temperature of $T_c = 287.7$ K at 267 GPa was reported. The upper critical magnetic field $H_{c2}(0) = 62$ T was found using measurements of $H_{c2}(T)$ up to 9 Tesla, and then applying Ginzburg–Landau theory. Also, using the equation

$$H_{c2}(0) = \frac{\Phi_0}{2\pi\xi(0)^2}, \quad (1)$$

the coherence length is determined to be $\xi(0) = 2.3$ nm.

If we now use the following expression [8] to obtain the London penetration depth at $T = 0$,

$$\lambda(0) = \frac{\Phi_0}{2\sqrt{2}\pi H_c(0)\xi(0)}, \quad (2)$$

and (incorrectly) insert the 62 T value of $H_{c2}(0)$ for $H_c(0)$, we would obtain $\lambda(0) = 1.6$ nm. According to H&M [7], the $\lambda(0) = 3.8$ nm value reported by Snider *et al.*[6] is claimed by the authors to be a typo. However, because $H_{c2}(0) \neq H_c(0)$, the 1.6 nm value is also incorrect. As H&M also point out, the experiment does not in fact provide the data to estimate $\lambda$.

However, it is possible to estimate $\lambda$ by using isotropic BCS theory [9] and an estimate of the Fermi velocity using the relationship [10]

$$\xi(0) = \frac{\hbar v_F}{\pi \Delta(0)}. \quad (3)$$

In order to estimate $\Delta(0)$, we need to use the $T_c$ reported by the experiment and the relation $2\Delta(0) = r k_B T_c$, where $r = 3.52$ in the weak coupling (BCS) limit, but is likely higher in this case, since in other high-pressure hydrides there is evidence for strong coupling. In particular, for both H$_3$S and LaH$_{10}$, the electron-phonon coupling constant is estimated to be [11] ~2 and $r$ is 4.5–4.8 for pressures around 200 GPa [11–13]. Also, based on the reported $R(T, H)$ curves for C-S-H, Talantsev has estimated the electron-phonon coupling constant to be ~2 as well [14]. We proceed with a modest strong coupling value of 4.5, and obtain $\Delta(0) = 56$ meV. This yields the Fermi velocity as

$$v_F = \frac{\pi \xi(0) \Delta(0)}{\hbar} = 6.1 \times 10^5 \text{ m/s}. \quad (4)$$

Because of the absence of a band structure calculation for this material, we use the free electron gas approximation to estimate other quantities. We discuss the validity of this approach below. From the relationship

$$v_F = \sqrt{\frac{\hbar^2}{m_e^2}(3\pi^2 n)^{2/3}} \quad (5)$$

the carrier density is estimated to be $n = 5.0 \times 10^{27}$ m$^{-3}$.

From this, the London penetration depth can be computed as

$$\lambda(0) = \sqrt{\frac{\varepsilon_0 m_e c^2}{n e^2}} = 75 \text{ nm}, \quad (6)$$

which yields the Ginzburg–Landau parameter as $\kappa = \lambda(0)/\xi(0) = 75 \text{ nm}/2.3 \text{ nm} = 33$, resulting in a strongly type II superconductor.

As pointed out by H&M, for strongly type II superconductors, we expect the *R vs. T* curves at the phase transition to be much less sharp than reported in the experiment. In Figure 2(b) of Ref. [6], the transition width ($\Delta T_c$) appears to be less than 1 K (see also Extended Data Figure 7(b)), yielding $\Delta T_c / T_c < 0.003$ for all the values of the applied magnetic field. In addition to the comparison made by H&M between the C-S-H system and MgB2, which has $\Delta T_c / T_c = 0.08$, let us compare it with the other two high-pressure hydrides with high $T_c$ [1–3]. By repeating the calculations above, using the $T_c$ and $H_{c2}(0)$ values provided by Refs. [1,3], *i.e.* $T_c = 185$ K and 73 T for H3S at 195 GPa, and $T_c = 250$ K and 115 T for LaH10 at 150 GPa, and setting $r = 4.5$, we find

$$\text{H}_3\text{S}: \quad \xi(0) = 2.12 \text{ nm}, \quad \lambda(0) = 164 \text{ nm}, \quad \kappa = 77;$$
$$\text{LaH}_{10}: \quad \xi(0) = 1.69 \text{ nm}, \quad \lambda(0) = 147 \text{ nm}, \quad \kappa = 87.$$

These values indicate that these two hydrides are also strongly type II like the C-S-H system. We note here that the hydride systems mentioned are expected to be far from defect-free single crystals, let alone be described by a free electron model [15]. However, it still provides a reasonable estimate for $\kappa$ and $\lambda(0)$: The magnetization measurements that are available for H3S yield [1] $H_{c1}(0) = 0.03$ T, which then allows us to estimate $\kappa \cong 59$ and $\lambda(0) \cong 125$ nm using the formula

$$\frac{H_{c1}(0)}{H_{c2}(0)} \cong \frac{\ln \kappa}{2^{3/2} \kappa^2} . \quad (7)$$

Additionally, Talantsev has estimated $\lambda(0)$ for H3S to be 189 nm [16]. Therefore, although it does not capture the underlying physics of the system, the free electron approximation offers a reasonable method of estimating the London penetration depth when one is limited to the experimental data which provides only $T_c$ and $H_{c2}(0)$. Thus, we are confident that the C-S-H system is strongly type II if it is a superconductor.

To complete the comparison, we get $\Delta T_c / T_c = 30/195 = 0.15$ for H3S using Figure 2(b) of Ref. [1], and $\Delta T_c / T_c = 15/245 = 0.06$ for LaH10 using Figure 1 of Ref. [3]. These values are 20 to 50 times greater than the reported value corresponding to Ref. [6]. Although there is no established universal theoretical lower limit for $\Delta T_c / T_c$, we are not aware of any systems other than simple elemental superconductors where this ratio is less than 0.01. Any inhomogeneities, thermal fluctuations, thermal gradients and grain/surface effects will cause the transition width to increase [17], most of which the C-S-H system should amply possess.

Additionally, in a type II superconductor, the transition width $\Delta T_c/T_c$ is predicted to increase with increasing applied magnetic field within multiple frameworks [17–19]. However, the reported behavior of the C-S-H system shows no such increase: comparison of the different curves in Figure 2(b) of Ref. [6] yield indistinguishable transition widths. Such a transition, which remains sharp for even the highest applied magnetic field, results in anomalous values for other physical quantities. For instance, the behavior of resistance can be fitted to Equation (6) of ref. [18], which, to give $\Delta T_c/T_c < 0.003$ at 9 Tesla, would require the fitting parameter $A$ to be greater than 600,000 Tesla and give a critical current density around $2 \times 10^9$ A cm$^{-2}$. Such a current would be two orders of magnitude larger than the highest-ever achieved current density to date [20]. To make the critical current density 2 orders of magnitude smaller, the $\Delta T_c/T_c$ ratio should be 0.04 or larger at 9 Tesla.

Therefore, either this C-S-H system is an anomalous superconductor that exhibits some heretofore unknown behavior, or the reported measurements arise from physical mechanisms not related to superconductivity. If it is the latter, then the observed sharp change in resistance with rising temperature could be due to a transition from a metallic state to a semimetallic or semiconducting state.

In order to estimate the resistivity of the C-S-H system above $T_c$, we take the reported resistance and sample size and follow the simple formula derived for the four-point van der Pauw procedure[21]

$$\rho \cong \frac{\pi d R}{\ln 2}, \quad (8)$$

where $d$ is the thickness of the sample ($5 - 10$ μm) and $R$ is the measured resistance ($1 - 2$ Ω) as given in Ref. [6]. This yields $\rho$ values in the range of $2 \times 10^{-5} - 9 \times 10^{-5}$ Ωm, which falls into the poor metal/semimetal range. Below $T_c$, $R$ drops by about 2–3 orders of magnitude [6], bringing $\rho$ into the range of typical metals.

**Discussion**

We have established that if the reported C-S-H is indeed a superconductor, it is an anomalous one. If it is the case that the below-$T_c$ state is not superconducting, we offer two alternative speculations for mechanisms that could result in the reported experimental data:

Case (i): A band structure transition that results from the change of the unit cell volume with changing temperature. In this scenario, the temperature sweep plays a role parallel to a pressure sweep. With increasing temperature, the unit cell expands, which might change the band overlaps in discontinuous ways. An abrupt transition at the direct band gap with changing pressure was recently observed for solid hydrogen [22], and discontinuous band structure changes that result in a transition between a metallic state and a semimetallic state were proposed as a possible explanation for this observation [23]. It is not clear, however, that this type of electronic mechanism would provide such a sharp jump in the resistivity measurements.

Case (ii): A structural phase transition from a metallic to semimetallic/semiconducting configuration. To understand this scenario, we can use VO$_2$ as a model, which has a structural phase transition [24,25] around 340 K accompanied by a sharp change in resistance ($\Delta T_c / T_c = 0.02$) [25]. Although it has been known for decades, the mechanism driving this transition is still not fully understood. One of the major proposals is that it is largely driven by anharmonic lattice dynamics [24]. Because anharmonic effects should be important in this high-pressure hydrogen-based system [23] and alternative crystal structures with nearly equal enthalpies might be available, an analogous phase transition is a possible scenario in the C-S-H system.

In both of the above proposals, the observed response of the phase transition to a magnetic field remains to be explained. Although this system should ordinarily not be magnetic, because the chemical formula and the crystal structure are not known, it is possible that the system has unpaired electrons that can align with an applied magnetic field, as has been predicted for amorphous carbon [26]. If that is the case, the shift of the transition temperature with applied magnetic field could be due to the magnetization in the system. Similarly, dependence of the metal-insulator transition temperature on an applied magnetic field has been established in other materials [27].

**Conclusion**

Based on well-established equations that describe a superconducting material, we have demonstrated that the *R vs. T* measurements of C-S-H reported by Snider *et al.* [6] deviate significantly from the expected behavior. We have also estimated the resistivity of the experimental sample, which is in the range of typical metals below $T_c$. In conclusion, if the newly discovered C-S-H is a superconductor, it shows anomalous behavior, and if it is not a superconductor, the measured data could be the result of a transition between a metallic state and a semimetallic/semiconducting state, possibly accompanied by a structural phase transition.

**Acknowledgments**

This work was supported primarily by the Director, Office of Science, Office of Basic Energy Sciences, Materials Sciences and Engineering Division, of the U.S. Department of Energy under contract No. DE-AC02-05-CH11231, within the Theory of Materials program (KC2301). Further support was provided by the NSF Grant No. DMR-1926004. We thank Yves Petroff and Warren Pickett for helpful comments.